\documentclass{mem}
\usepackage{txfonts}
\usepackage{balance}
\usepackage{graphicx}
\usepackage{enumitem}
\usepackage{amssymb}
\usepackage{wrapfig}

\idline{75}{282}
\begin{document}
\def\teff{$T\rm_{eff }$}
\def\kms{$\mathrm {km s}^{-1}$}

\title{Synergies of THESEUS with the SKA: a brief report}
%   \subtitle{}

\author{S. Colafrancesco \inst{1}}

\institute{
School of Physics, 
University of Witwatersrand, Johannesburg, RSA\\
\email{sergio.colafrancesco@wits.ac.za}
}

\authorrunning{S. Colafrancesco}

\titlerunning{Synergies with the SKA}

\abstract{We present a short report on the main synergies between Theseus and SKA in the study of high-redshift transients and we summarize a few more aspects where Theseus and SKA can contribute to explore fundamental physics in the universe.

}
\maketitle{}

\section{Theseus and SKA}

The Transient High Energy Sky and Early Universe Surveyor (THESEUS) is a space mission concept 
whose core science goals of THESEUS comprise the exploration of the cosmic dawn and re-ionization era through the prompt detection and characterization of  transients  up to high redshifts.\\
The Square Kilometer Array (SKA) has similar key science programs working in the very-low (70-350 MHz) to mid- (0.4-10 GHz) radio frequency range (for a comprehensive illustration of the SKA science topics and technological advancements see "Advancing Astrophysics with the Square Kilometre Array, 2014").

We can use (fast) transient source studies to perform population studies (blob-ology) of these kind of sources with their cosmological tests, studies of the time-domain universe (pop-ology) using the fast and short transient sources that we can observe out to high redshifts, but we can also explore some fundamental aspects of the physics of the universe related to these cosmic explosions.

\subsection{Blob-ology}

This is mainly the case of the GRB (radio afterglows) population study.
The first moral we can get from this study is that the sensitivity of the SKA will allow to observe almost the complete population of GRBs, provided that a X/g-ray instrument (more sensitive than Swift) will be operational in the SKA era (e.g., Theseus)  and provided a GRB localisation to a few arcsec resolution \cite{Burlon2015}.

The second moral we can get  is that once you have a relatively large sample of sparks (GRBs) over a large area of the sky with spectroscopic identification, then you can probe the geometry of the Universe
that depends on its fundamental matter-energy constituents  \cite{Amati2014}.

\subsection{Pop-ology}

The moral we can get in this field is that impulsive radio bursts that are detectable across cosmological distances are extremely powerful probes of the ionized Inter-Galactic Medium (IGM), intergalactic B-fields, and the properties of space-time  \cite{Macquart2015}.

The synergies between SKA and Theseus can help exploring a vast part of the transient parameter space of specific luminosity vs. the product of observing frequency and transient duration, thus allowing to extract basically all the relevant information on a wide range of fast and short transient sources of coherent emission.

\subsection{Physics}

Among the fundamental physics questions that a strong synergy between SKA and Theseus can explore there are: i) the physics of GW event counterparts, ii) LIV tests and ALP physics from AGN sparks, iii) the fundamental nature of the photon, iv) the physics of PBHs, v) the exploration of Quantum Gravity effects through Planck Stars.

As an example of the fruitful SKA-Theseus synergy I want to mention here the case of Planck Stars \cite{Rovelli2017}. Primordial BHs can decay over a time scale 
$$
\tau \approx {m^2 \over m^{2}_{Planck}} \tau_{Planck}
$$
When they explode at a later stage than the Planck time, they emit two typical kind of signals:\\
i) a low-E component of the signal emitted at the explosion should be a powerful burst with wavelength of the order of the size of the hole, e.g.: mm. to cm.;\\
ii) a high-E component is related to the photon gas originally collapsed in the early universe, and then liberated after the short-internal/long-external time \cite{RovelliVidotto2014}. 

If Planck Stars (PS) have an initial mass distribution, then they also have a distribution of exploding times, with lower size PS exploding at high-z and higher size PS exploding at low-z.
This fact enhances the rate of remote Planck Stars at high-z.\\
It turns out that  high-z PS are more visible in the X-rays / soft gamma-rays by Theseus, while
low-z  PS are visible in the radio-mm frequency range with the SKA \cite{Colafrancesco2018}.

This is only one example of the tremendous potential of studying the fundamental physics at work in the high-z and low-z universe using the possible synergy between SKA and Theseus.
These two projects will possibly become operative in the next decade with enormous potential in cross-correlated studies of transients in the Universe.

%
%\bibliographystyle{aa}
%\bibliography{sxi}

\begin{thebibliography}{9}

\bibitem{Amati2014} Amati, L. 2014, Annalen der Physik, Vol. 526, issue 7-8, p.340-346

\bibitem{Burlon2015} Burlon, D. et al. 2015, Proceedings of Advancing Astrophysics with the Square Kilometre Array (AASKA14). 9 -13 June, 2014. Giardini Naxos, Italy. Online at http://pos.sissa.it/cgi-bin/reader/conf.cgi?confid=215, id.52

\bibitem{Colafrancesco2018} Colafrancesco, S. et al. 2018: in preparation 

\bibitem{Macquart2015} Macquart, J.P. et al. 2015, Proceedings of Advancing Astrophysics with the Square Kilometre Array (AASKA14). 9 -13 June, 2014. Giardini Naxos, Italy. Online at http://pos.sissa.it/cgi-bin/reader/conf.cgi?confid=215, id.55

\bibitem{Rovelli2017} Rovelli, C. 2017, Nature Astronomy, Volume 1, id. 0065

\bibitem{RovelliVidotto2014} Rovelli, C. \& Vidotto, F. 2014, International Journal of Modern Physics D, Volume 23, Issue 12, id. 1442026

\end{thebibliography}

\end{document}